\documentclass[aps,prfluids,titlepage,showpacs,a4paper,preprint,longbibliography,
tightenlines,superscriptaddress,nofootinbib]{revtex4-2}
\usepackage{graphicx,xcolor}
\usepackage{amsmath}
\bibpunct{(}{)}{;}{a}{ }{,}

\usepackage{epsfig,latexsym,array}
\newcommand{\be}{\begin{equation}}
\newcommand{\ee}{\end{equation}}
\newcommand{\ba}{\begin{eqnarray}}
\newcommand{\ea}{\end{eqnarray}}
\newcommand{\benn}{\begin{displaymath}}
\newcommand{\eenn}{\end{displaymath}}
\renewcommand{\d}[1]{{\rm d}#1}

\renewcommand{\vec}[1]{\mbox{\boldmath $#1$}}

\newcommand{\e}[1]{\mbox{e}^{#1}}
\newcommand{\wt}[1]{\widetilde{#1}}

\newcommand{\kwave}{k}
\newcommand{\flucu}{u} 
\newcommand{\flucv}{v} 
\newcommand{\flucw}{w} 
\begin{document}
\title[Linear response of turbulence]{Linear response of turbulence}
\author{H. E. Cekli and G. Bertens and W van de Water}
\affiliation{Physics Department, Eindhoven University of Technology\\
5600 MB Eindhoven, The Netherlands\\ International Center for
Turbulence Research} 
\email[Corresponding author: ]{w.vandewater@tudelft.nl}
%
\date{5 august 2025}  
\begin{abstract}
We study the response of wind tunnel turbulence to perturbations
using an active grid.  We compare our findings to Kraichnan's linear
response result $R(\kwave,\tau_d) = \exp(-\kwave^2 \: \tau_d^2 \:
\flucu^2)$ which predicts a decay of the response with increasing
turbulence intensity $\flucu$, wave number of the perturbation
$\kwave$ and delay time $\tau_d$ since the perturbation was applied
\citep{kraichnan.1964}.
In our experiments we used two different mechanisms to create a
perturbation of a pre--existing and well--developed turbulent flow.
In the first case we combine both the turbulence generation and the
additional random perturbation in the same active grid.  We find that
the reponse decays with increasing $\tau_d$, but much slower than
predicted, while the decay was virtually independent of the
turbulence intensity.
In the second type of experiments, we perturb an active
grid--generated turbulent flow using a loudspeaker--driven synthetic
jet.  The perturbation is at a single wave number $\kwave$, placed
well inside the inertial range.
The response was found to decay with increasing time delay $\tau_d$,
while the decay is faster for larger wave numbers, roughly as
predicted by Kraichnan's model.  However, also in this case the decay
rate is independent of the turbulence intensity.
\end{abstract}
\maketitle
%
%
\section{Introduction} 
The fluctuation-dissipation theorem of statistical physics expresses
that in thermal equilibrium the linear response of a system is
proportional to the magnitude of its intrinsic thermal fluctuations.
It is tantalizing to apply this concept to turbulence, as it would
link the response of a turbulent flow to forcing to the size of its
spontaneous velocity fluctuations. The problem is that turbulence is
not linear and it is not in thermal equilibrium. Nevertheless, the
idea of linear response is central to Kraichnan's direct interaction
approximation, and there have been several experimental and numerical
attempts to measure the linear response of turbulence. 

Understanding the response of turbulence to perturbation is the
central issue in the field of turbulence control. This is an
important technological application which aims at reducing turbulent
drag or delay separation. 
A large body of work exists where the methods of linear control
theory are aplied to the turbulent boundary layer \citep{kim.2003}.

The linear regime dictates very small perturbation magnitudes, which
presents an experimental and numerical challenge.  
A more philosophical problem is the definition of the response
function in an experiment. In numerical simulations, the unperturbed
state can be specified completely in a perfectly reproducible way.
Averages can then be performed over an ensemble of unperturbed
states. This is impossible in an experiment where averages have to be
done over the perturbed state instead.

The response of a laminar boundary layer was first studied by
\cite{schubauer.1947} and that of a turbulent boundary layer by
\cite{hussain.1970}.  In both experiments the perturbation was at a
single wave number and frequency.  
The response of homogeneous grid--generated turbulence was measured
by \cite{kellogg.1980} and \cite{itsweire.1984}.
The experimental results were compared to Kraichan's simple advection
model \citep{kraichnan.1964} in which the forced small scales are
advected randomly by the large scales, for which Gaussian statistics
is assumed.  This model provides an attractive organizing principle
for experiments and simulations, with the linear response function:
\begin{equation} 
	R(\kwave,\tau_d) = 
    e^{-\frac{1}{2} \: \kwave^2 \: \tau_d^2 \: \flucu^2},
\label{eq.dia}    
\end{equation}
which decays very rapidly with increasing large--scale turbulence
intensity $\flucu$, increasing wave number $\kwave$ of the
perturbation, and increasing elapsed time $\tau_d$ since the
perturbation was applied.  
In order to appreciate the response function (Eq.\ \ref{eq.dia}), let
us briefly sketch a derivation \citep{kraichnan.1964}.  In the random
advection problem, the small--scale velocity field $v$ is advected by
a large--scale Gaussian field $u$ and forced by $F$, there is no
other dynamics.  Thus, ignoring pressure and viscosity, the Fourier
amplitudes $\wt{v}_i$ satisfy
\benn
   \frac{\partial \wt{v}_i}{\partial t} - \imath k_j\: u_j \wt{v}_i =
   \wt{F}_i.
\eenn
If the forcing is delta--correlated in time, this equation yields the
linear response (Eq.\ \ref{eq.dia})
\benn   
   \left\langle \wt{v}_i(\vec{k}, t + \tau_d) \:
   \wt{F}_i^*(\vec{k}, t) \right\rangle = 
   \e{-\frac{1}{2} \: k^2 \: \tau_d^2 \: \flucu^2}
   \left\langle \wt{F}_i(\vec{k}, t) \: \wt{F}_i^*(\vec{k}, t)
   \right\rangle.
\eenn   
Central to the derivation of (Eq.\ \ref{eq.dia}) is a separation of
the scales of the fluctuations $v$ and the force, and those of the
random field $u$.  Therefore, the force has to act at inertial--range
scales.
An experimental attempt to measure the response of homogeneous
turbulence was made by \cite{kellogg.1980}. 
They disturbed a standard grid--generated, decaying and
nearly--isotropic turbulent flow in a wind tunnel using a separate
extra grid to impose the perturbation. The first grid produced the
turbulence while the second grid, which consisted of a
one-dimensional mesh of parallel wires, imposed the perturbation at
wave number $\kwave$, with $\kwave$ determined by the separation of
the wires.  Similar experiments were done by \cite{itsweire.1984}.
Both articles document the distortion of the spectrum due to the
perturbation. \cite{kellogg.1980} found a near--perfect exponential
decay with distance of the oscillation amplitude in the two--point
correlation function.  
In these experiments, the time delay $\tau_d$ since application of
the perturbation is set by the separation $x_1$ to the second grid as
$\tau_d = x_1 / U$, where $U$ is the mean velocity. 

Surprisingly, although (Eq.\ \ref{eq.dia}) implies an exponential
decay with increasing $\tau_d^2$, rather than with increasing
$\tau_d$, \cite{kellogg.1980} conclude good agreement of their
response function with (Eq.\ \ref{eq.dia}). Kellog and Corrsin also
consider a time--dependent perturbation using an oscillating wire and
suggest modulation using a kind of active grid.  No response function
for these alternative excitations were reported.  The present paper
was inspired by their suggestion.

\cite{camussi.19971} studied the evolution of a perturbation
generated by a small pulsed jet in the presence of a turbulent jet
flow. They measured the response function as the Fourier transform of
the velocity, conditionally averaged on the pulsed perturbation.
Agreement with (Eq.\ \ref{eq.dia}) was found. Similarly to the
present experiment, the time delay was set by the spatial separation
$x_1$ of the probe and perturbation. However, with $x_1$ also the
turbulence intensity changed, so that $\tau_d$ and $\flucu$ varied
simultaneously in these experiments.

Evidence for the response function (Eq.\ \ref{eq.dia}) was found in
direct numerical simulations of turbulence by \cite{carini.2009}.  By
forcing with white noise, the response function is computed
simultaneously at all wave numbers.  A similar appoach was used by
\cite{luchini.2006} to compute the response of a turbulent channel
flow.
Using the axisymmetry of a turbulent von K\'arm\'an flow,
\cite{monchaux.2008} identified fluctuation--dissipation relations in
their analysis of the fluctuating velocity fields measured in their
experiment.

Let us now sketch how we will confront (Eq.\ \ref{eq.dia}) with
experiments. Our experiments are done in a recirculating wind tunnel,
in which relatively small turbulent velocity fluctuations
$(\flucu\approx 1 \: \text{ms}^{-1})$ are carried by a mean wind
$(U\approx10 \: \text{ ms}^{-1})$. We will impose perturbations on
this turbulent flow as temporal modulations at a fixed location, and
study their fate at various downstream locations, with $x_1$ the
distance to the source of the perturbation. Through Taylor's frozen
turbulence hypothesis, these temporal modulations, done at frequency
$\omega$, translate to spatial modulation with wave number $\kwave =
\omega/U$. For large separations $x_1$, corresponding to times
$\tau_d = x_1 / U$ much larger than $2 \pi / \omega$, we see the
temporal decay of the modulations. 
Consequently, the delay time $\tau_d$ since the application of the
modulation was varied by measuring the response at various
separations from the perturber. 

With a (fixed) conventional grid, the mean velocity determines the
turbulent fluctuations $\flucu$. With an active grid, both quantities
can be varied independently, which allows us to keep time delay
$\tau_d$ constant while varying the the turbulence level.

Guided by the expression for the response function (Eq.\
\ref{eq.dia}),
we will first determine the niche in the parameter space of our
experiment where we should study the response function.  Implicit in
(Eq.\ \ref{eq.dia}) is the separation of large and small scales, so
that the wave number of the perturbation should be in the inertial
range. The wave number is determined by the forcing frequency $f_p$
and the mean velocity $U$ as $\kwave = 2 \pi f_p/U$.  The inertial
range starts at the large--eddy turnover rate $u / M$, with $M$ the
grid mesh size, so that $\kwave \gtrsim 2\pi u / (M U)$. 
In our experiment, a typical mean velocity is $U = 10 \text{
ms}^{-1}$, a typical fluctuating velocity is $\flucu = 1 \text{
ms}^{-1}$, and a typical separation between perturbation and
detection is 4~m, so that
$\exp(-\kwave^2 \: \tau_d^2 \: \flucu^2 / 2) \approx 4 \times
10^{-2}$,
which should be measurable in an experiment. However, the niche in
parameter space is very small. For example, doubling the wave number
decreases the response by more than two orders of magnitude.
It must therefore be compensated by moving closer to the source of
the perturbation.

The challenge of the experiment is to devise a perturbation which is 
independent of the background turbulence.
This is a delicate issue, which also haunted the experiments of
\cite{kellogg.1980} and \cite{itsweire.1984}.  
The wind tunnel and the active grid will be described in \S\
\ref{sec.exp.set}.  Following a suggestion by \cite{kellogg.1980}, we
will first measure the response using random motion of the active
grid.  In \S\ \ref{sec.active.per} the modulation of turbulence will
be done by a small--amplitude random rotation of one axis, while the
other axes control the background turbulence level.  To make a broad
excitation spectrum, the motion of the modulated axis has a short
correlation time.
Perturbation experiments at much larger wave number which involve a
synthetic jet as the source of perturbations will be discussed in \S\
\ref{sec.synth.per}.

\section{Experimental setup}
\label{sec.exp.set}
Active grids, such as the one used in our experiment, were pioneered
by \cite{makita.1991} and consist of a grid of rods with attached
vanes that can be rotated by servo motors. The properties of actively
stirred turbulence were further investigated by 
\cite{mydlarski.1996} and \cite{poorte.2002}. 
Active grids are ideally suited to modulate turbulence in space-time
and offer the exciting possibility to tailor turbulence properties by
a judicious choice of the space-time stirring protocol
\citep{cekli.2010}. In our case, the control of the grid's axes is
such that we can prescribe the instantaneous angle of each axis
through a computer program. 
Our active grid has mesh size $M = 0.1\:{\rm m}$.  From the seminal
study of static grids by \cite{comte-bellot.1966} it follows that it
typically takes a downstream separation of $40 M$ for the flow to
become approximately homogeneous and isotropic.
%

\begin{figure}
\centerline{\includegraphics[width=13.0 cm,angle=0]{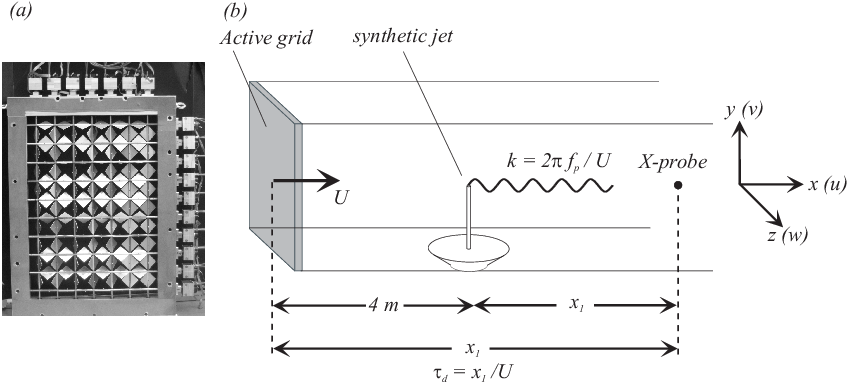}}
%
\caption{ 
(a) Photograph of the active grid. 
(b) Schematic view of the experimental setup. The synthetic jet is
absent in the experiments described in \S\ \ref{sec.active.per} where
the modulation is done with help of the active grid. The experiments
involving the synthetic jet are described in \S\ \ref{sec.synth.per}.
The jet is placed at a distance of 4~m (40 mesh sizes $M$) behind the
grid.  The delay time $\tau_d = x_1 / U$, with $x_1$ the distance of
the x--wire to the grid in \S\ \ref{sec.active.per}, and distance of
the x--wire to the jet orifice in \S\ \ref{sec.synth.per}.
}
\label{fig.geometry}
\end{figure}

The active grid is placed in the $8$~m--long experimental section of
a recirculating wind tunnel. Turbulent velocity fluctuations are
measured using a two--component hot--wire (x--wire) anemometer. The
locally manufactured hot--wires had a $2.5$ $\mu$m diameter and a
sensitive length of $400$ $\mu$m and were operated at constant
temperature using computer--controlled anemometers that were also
developed locally. Each experiment was preceded by a calibration
procedure.  The x--wire probe was calibrated using the full velocity
versus yaw angle approach \citep{browne.1989,tropea.2007}. The
resulting calibrations were updated regularly during the run to allow
for a (small) temperature increase of the air in the wind tunnel. The
signal captured by the sensor was sampled simultaneously at $20$ kHz,
after being low--pass filtered at $10$ kHz.  
The mean $U = \langle u(t) \rangle_t$ and turbulent $u = \langle
(u(t) - U)^2 \rangle_t^{1/2}$ velocities, and similarly for $v, w$,
are defined as time averages.

The grid is operated by a computer, and the instantaneous angle of
each rod is recorded to define the grid state, which can be
correlated with the measured instantaneous velocity signal. In Fig.\
\ref{fig.geometry} a photograph of the grid is shown, together with a
sketch of our experiment geometry. 

\section{Active--grid perturbations}
\label{sec.active.per}
In our experiments we use the same active grid, not only to generate
a turbulent velocity field in the wind tunnel, but also to impose
perturbations on top of the generated turbulence. 
A challenge is the design of grid protocols where the generation of
perturbations and the stirring of background turbulence are decoupled
as well as possible.  

\begin{figure}
\centerline{\includegraphics[width=7.8cm]{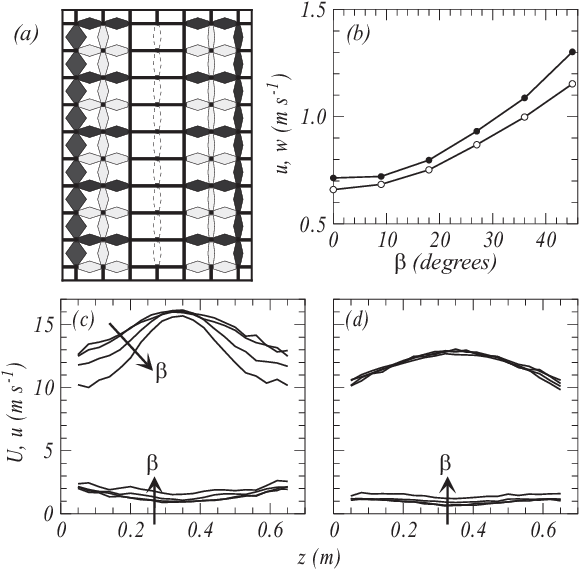}}
\caption{ 
(a) Modified active grid for the experiments of \S\
\ref{sec.active.per} in which both the background turbulence and its
modulation were done by the active grid.  The central axis is used
for the perturbation. The two center rows of blades on the horizontal
axes have been removed and the two vertical axes next to the central
one are kept stationary. 
(b) Controlling the background turbulence level through the grid
angle $\beta$; the velocities $u$ (closed dots) and $w$ (open
circles) have been measured at a downstream location of $x_1 = 4.6
\:{\rm m} \; (= 46 \: M)$ of the active grid.
(c, d) Profiles of mean and turbulent velocities $U, \flucu$ at $x_1
= 1.6\: {\rm m}$ (c), and $x_1 = 4.6\: {\rm m}$ (d). The profiles are
shown for grid angles $\beta = 0, 9, 27$ and $45\:^\circ$. The
central (modulation) axis is kept stationary. 
}
\label{fig.pertur_grid_all}
\label{fig.U_rms_beta}
\end{figure}

To this aim we modified the active grid as illustrated in Fig.\
\ref{fig.pertur_grid_all}(a).  The turbulent flow generated by the
grid consists of two turbulent wakes from the partially stationary,
partially moving sides of the grid while an isolated axis in the
center provides the modulation. 
This modulation is done by rotating the axis randomly within a given
range around a mean angle of zero (blades in the stream--wise
direction).
The time--dependent angle is $\alpha(t) = \Delta \: \theta(t)$ with
maximum amplitude $\Delta=7.2^\circ$.  The time series $\theta(t)$
consists of uncorrelated random numbers, uniformly distributed on the
interval $\left[-{1}/{2},{1}/{2} \right]$ and sampled at 20~Hz.
In our experiment it is not possible to specify the forcing
$\vec{F}(\vec{x},t)$ which corresponds to a wind which passed
through an active grid. 
Therefore we now simply assume that ${F}$ is proportional to the
instantaneous angle of this rod.  For computation of the correlation
with the wind, the axis angle is sampled at 500~Hz.
The neighboring vanes of the perturbating axis on the horizontal rods
have been removed and adjacent vertical rods on either side are kept
stationary at $0^\circ$ to isolate the perturbation from the
generation of the background turbulence. Finally, two vertical rods
at each border are rotating randomly to improve homogeneity. 

The horizontal axes are used to generate the background turbulence.
We tune the turbulence intensity by setting these axes to a fixed
angle $\beta$ in an alternating fashion; the dependence of $\flucu,
\flucw$ on $\beta$ is illustrated in Fig.\ \ref{fig.U_rms_beta}(b).
The change in the mean velocity due to the change of the grid
transparency was compensated by regulating the wind--tunnel fan
power, such that the mean velocity (which determines the time delay
$\tau_d$) was kept at $U = 13.4 \text{ ms}^{-1}$.

The wake of such a grid is not homogeneous; as Fig.\
\ref{fig.U_rms_beta} illustrates, the mean velocity is largest in the
center, where the fluctuating velocity is smallest. Further
downstream at $x_1 = 4.6\: {\rm m}$, the mean velocity profile
becomes independent of $\beta$, while the turbulent velocity
uniformly increases with $\beta$.

In our experiments we measure the cross--correlation between the
modulation axis angle $\alpha$ and the $z-$component of the velocity
$w$ which carries the largest imprint of the perturbation.
Long averaging times were needed to detect the effect of the
perturbation in the turbulent signal; the shown results were averaged
over $10^3$~s corresponding to $2.5\times 10^3$ large-eddy turnover
times. 
The experiments were repeated at different downstream locations
behind the active grid ranging from $x_1 = 1.15$ to $6.55$~m, with
the associated time delays $\tau_d = x_1 / U$, and as a function of
the background turbulence levels.

We define the correlation function in the usual way,
\be
   C_{wF}(\tau) = \frac{\langle F(t) w(t+\tau)  \rangle - 
   \langle F(t) \rangle \langle w(t) \rangle}
   {\left ( \langle F(t)^2 \rangle - 
   \langle F(t) \rangle^2 \right)^{1/2}  
   \left( \langle w(t)^2\rangle -  \langle w(t) \rangle^2
   \right)^{1/2}},
\label{eq.corrfun}
\ee
where the force $F$ is taken to be the axis angle $\alpha$ and
averages are done over time.
%

\begin{figure}
\centerline{\includegraphics[width=13.8 cm,angle=0]{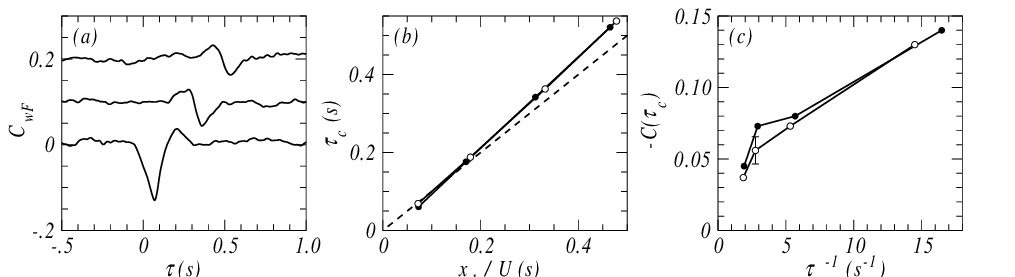}}
%
\caption{ 
  (a) Correlation $C_{w F}$ between the perturbations and measured
  turbulent velocity at $x_1 = 1.2, 4.6$ and $6.6\:{\rm m}$
  downstream from the active grid.  The curves at $x_1 = 4.6$ and 
  $6.6\:{\rm m}$ have been shifted vertically by 0.1 and 0.2,
  respectively. The correlation functions have a sharp dip at
  $\tau_c$, which is the convection time of the perturbation with the
  mean flow. 
  (b) Convection time $\tau_c$ as a function of the distance $x_1$ to
  the grid. Dashed line: $\tau_c = x_1 / U$.
  (c) $-C_{w F}(\tau_c)$ as a function of time delay $\tau_c$.  The
  magnitude of the correlation does not depend significantly on the
  turbulent velocity $u$, the error bar indicates the variation for
  $u$ ranging from 0.72 to $1.30\; {\rm m}{\rm s}^{-1}$.  Open
  circles: constant angle $\beta$ of the turbulence generator, closed
  dots: $\beta$ adjusted so that at each separation the background
  turbulence level is the same.
} 
\label{fig.fd4_correlation}
\end{figure}

The normalized correlation function (Eq.\ \ref{eq.corrfun}) between
the perturbation and the velocity signal $C_{wF}(\tau)$ is shown in
Fig.\ \ref{fig.fd4_correlation}(a) at several separations $x_1$.  The
time $\tau$ at which the sharp dip occurs is the convective time
$\tau_c$ of the perturbation by the mean flow; it is shown as a
function of $x_1$ in Fig.\ \ref{fig.fd4_correlation}(b).  Because the
shape of the correlation function evolves with $x_1$, the dip lags
significantly behind the mean flow.

The magnitude $|C_{wF}(\tau_c)|$ is shown in Fig.\
\ref{fig.fd4_correlation}(b) as a function of the delay time
$\tau_c$. According to the response function (Eq.\ \ref{eq.dia}), for
a white--noise forcing $F$, $|C_{wF}(\tau_c)| \propto 1 / u^2
\tau_c$. The $\tau_c^{-1}$ time dependence holds approximately, but
the linear extrapolation of the correlation at the smallest delays
does not vanish at large times, and a long--time correlation remains.
In contrast to the prediction of (Eq.\ \ref{eq.dia}), the decay
does not depend significantly on the turbulent velocity $u$.  

In conclusion, a random modulation of a turbulent flow can still be
detected at large separations from its source.  However, its decay is
slower than predicted by (Eq.\ \ref{eq.dia}), while it does not
depend significantly on the background turbulence.  We finally notice
that the perturbation is small; measured velocity spectra with and
without the perturbing vanes moving were indistinguishable.

\section{Synthetic--jet perturbations}
\label{sec.synth.per}
Using an active grid whose driving frequency is limited by mechanical
constraints, it is very difficult to probe wave numbers deep in the
inertial range.  
Higher frequencies could be reached using a synthetic jet which was
placed at a distance of 4~m from the active grid (see Fig.\
\ref{fig.geometry}).  The background turbulence is mainly controlled
by the active grid. As in the experiments of \S\
\ref{sec.active.per}, the angles $\pm \beta$ of the stationary
horizontal rods control the background turbulence level, while the
vertical rods rotate randomly.
The used perturbation frequencies were $f_p = 40$~Hz and $70$~Hz,
corresponding to wave numbers $\kwave_p = 2 \pi f_p / U = 33.5$ and
$58.6 \text{ m}^{-1}$, respectively. The mean velocity in these
experiments is $U = 7.5 \text{ ms}^{-1}$. The velocity field is
measured by an x-probe at downstream locations in the range between
$x_1 = 0.15 - 1.4$~m of the synthetic jet. The corresponding time
delay for this range is $\tau_d = 0.01 - 0.1$ s. Since the
perturbation wave numbers are much larger than in the experiments of
\S\ \ref{sec.active.per}, the response function decreases more
rapidly with increasing separation to the jet, but we estimate that
at $x_1 = 1.34$~m, the response should still be detectable. 

The synthetic jet is driven by a loudspeaker which is connected to a
tube.  The loudspeaker is attached to a $2$~cm thick PVC plate which
contains a cavity at the loudspeaker side such as to provide space
for the moving cone. A $50$~cm long tube with inner diameter $d =
1$~cm is placed in the center of the plate. At the loudspeaker end
this tube is perforated by a ring of $5$ mm diameter holes, so that
air can be entrained on the reverse cycle and the perturbation is
more effective. 
The properties of synthetic jets have been documented by
\cite{smith.1998} and \cite{glezer.2002}. 

The synthetic jet is flush mounted on the bottom surface of the wind
tunnel as illustrated in Fig.\ \ref{fig.geometry}(b), which minimizes
the additional disturbance of the synthetic jet on the flow. However,
the flow is obstructed by the tube in a way that will be documented
below. An amplified sinusoidal signal with desired frequency is used
to drive the loudspeaker; this signal is taken as the force $F(t)$.
The signal is sampled parallel with the velocity probe at $20$ kHz,
so that these two signals can be correlated.  In contrast to the
random perturbations in \S\ \ref{sec.active.per}, we now work at a
single frequency $f_p$.

\begin{figure}
\centerline{
\includegraphics[width=13 cm,angle=0]{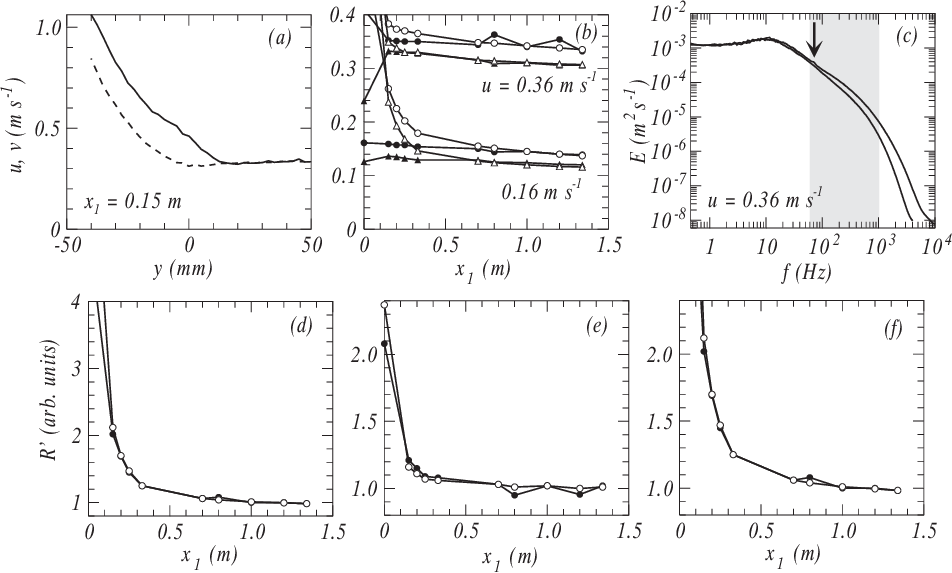}}
%
\caption{
(a) Measured $\flucv$ velocity profiles in the wind tunnel at
$15$~cm$ = 15 d$ of the synthetic jet, with $d$ the jet diameter. The
jet exit is at $y = -10\;{\rm cm}$.  Full line, with open synthetic
jet, dashed line: with synthetic jet blocked.
(b) Profiles of $u$ (circles) and $v$ (triangles) turbulent
velocities with closed and open synthetic jet for closed dots and
open symbols, respectively at background turbulent settings $u = 0.36
\: {\rm m s}^{-1}$ and $u = 0.16 \: {\rm m s}^{-1}$.  
(c) Spectra $E_{vv}(f)$, with closed and open synthetic jet at
forcing frequency $f_p = 70$~Hz (indicated by the arrow), $x_1 =
0.15$~m, and $u = 0.36\:{\rm ms}^{-1}$.  The integral over the shaded
frequency interval defines the quantity $R'$ (Eq.\ \ref{eq.rprime})
with $f_l = 30$~Hz and $f_u = 10^3$~Hz when $f_p = 70$~Hz, and
$f_l, f_u = 10, 1200$~Hz for $f_p = 40$~Hz.
(d, e, f) 
$R'$ for $f_p = 40\: {\rm Hz}, \flucu = 0.36\: {\rm m s}^{-1}$; 
$f_p = 70\: {\rm Hz}, \flucu = 0.36\: {\rm m s}^{-1}$; 
$f_p = 70\: {\rm Hz}, \flucu = 0.16\: {\rm m s}^{-1}$, respectively.
}  
\label{fig.perturb.jet}
\end{figure}

\begin{figure}
\centerline{\includegraphics[width=14 cm, angle = 0]{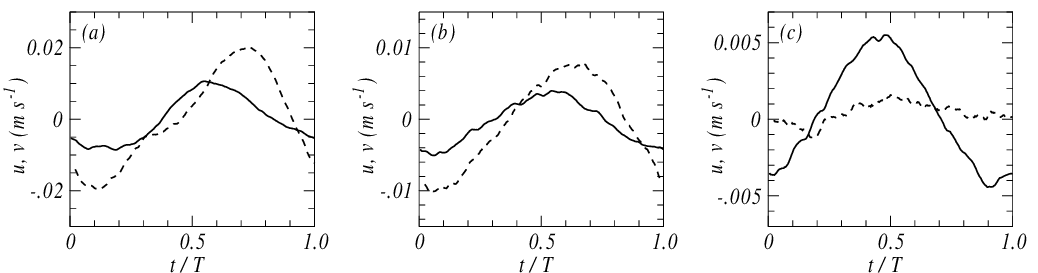}}
%
\caption{
Phase-averaged velocity in wind ($x-$) (full line) and jet exit
($y-$) (dashed line) directions measured at $x_1 = 0.2, 0.33, \: {\rm
and} \: 1.34\: {\rm m}$, for (a), (b), and (c), respectively.  The
forcing frequency is $f_p = 70\: {\rm Hz}$ (period $T = 0.014\: {\rm
s}$) while the background turbulent velocity is $\flucu = 0.36 \:
{\rm m s}^{-1}$. 
}
\label{fig.phase}
\end{figure}

A point of concern is the change of the background turbulence
properties induced by the perturbation.  Figure\
\ref{fig.perturb.jet}(a) shows the vertical profiles of the turbulent
velocity $\flucv$ with open and blocked jet.  The synthetic jet
injects air at height $y = -10 $~mm; except for Fig.\
\ref{fig.perturb.jet}(a), all other data are taken at $y = 0$. There
is a considerable variation in the profiles even when the synthetic
jet is passive, which is due to the wake created by the tube. 
When the synthetic jet is active there is a substantial increase in
the turbulent velocity which decays with increasing separation to the
jet, as is illustrated for two background turbulent intensities in
Fig.\ \ref{fig.perturb.jet}(b).  Especially for $u = 0.36\:{\rm m
s}^{-1}$, the turbulent intensity remains constant over the range of
separations considered, and the jet perturbation evolves through a
constant background turbulence.

In order to document the change of the spectrum when the jet is
driven, we integrate the spectra over a band of frequencies $f \in
[f_l, f_u]$, which includes the driving frequency $f_p$.  If
$E^o_{vv}(f)$ is the spectrum registered with the jet open, and
$E^c_{vv}(f)$ that with the blocked jet, we define the perturbation
as
\be
   R' = \left(\int_{f_l}^{f_u} E^o_{vv}(f) \: \d f
   \Big/ 
        \int_{f_l}^{f_u} E^c_{vv}(f) \: \d f \right)^{1/2}.
\label{eq.rprime}
\ee
The quantity $R'$, which merely quantifies the spectrum distortion,
cannot be compared to the response $R$ (Eq.\ \ref{eq.dia}) which,
unlike $R'$, vanishes if no phase relation exists between the forcing
and the response.

As Fig.\ \ref{fig.perturb.jet}(c) illustrates, the perturbation done
at a single wave number contaminates the spectrum at larger
frequencies. The influence on the background turbulence is shown in 
Fig.\ \ref{fig.perturb.jet}(d) for $f_p = 40\: {\rm Hz}, \flucu
= 0.36\: {\rm m s}^{-1}$, 
in Fig.\ \ref{fig.perturb.jet}(e) for $f_p = 70\: {\rm Hz}, \flucu =
0.36\: {\rm m s}^{-1}$, 
and in Fig.\ \ref{fig.perturb.jet}(f) for $f_p = 70\: {\rm Hz},
\flucu = 0.16\: {\rm m s}^{-1}$.
The addition of energy to the turbulent flow by the perturbation is
inevitable, as the perturbation must remain detectable.  The question
is how large it can be.  From Fig.\ \ref{fig.perturb.jet}(c) is
appears that the the change of the spectrum is mainly at wave numbers
larger than those of the perturbation.  These wave numbers do not
play a role in Kraichnan's random advection model (Eq.\
\ref{eq.dia}).

The evolution of the phase--averaged mean velocity with separation to
the source of the perturbation is shown in Fig.\ \ref{fig.phase}(a).
It is nearly sinusoidal, which suggests that the perturbed velocity
depends approximately linearly on the driving.  
Close to the jet, the phase average of the vertical velocity
component $v$ is largest, far away that of the horizontal
$u$--component.

\begin{figure}
\centerline{\includegraphics[width = 12 cm,angle=0]{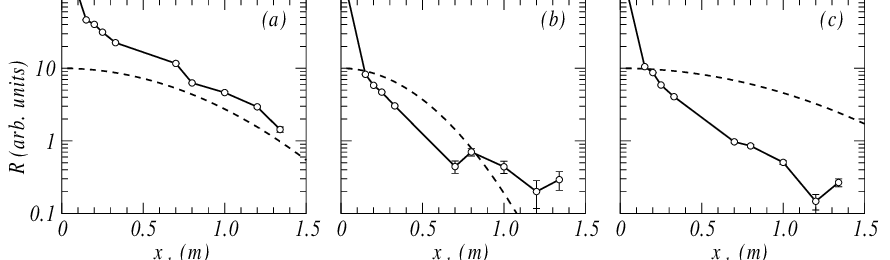}}
\caption{ 
  The spectral response $R=\left|\langle F^*(\kwave_p)\:w(\kwave_p)
  \rangle\right| $ compared with Kraichnan's prediction
  (Eq.\ \ref{eq.dia}) which is indicated by the dashed lines.
  (a) $\flucu=0.36 \text{ ms}^{-1}$, $f_p = 40$~Hz, 
  ($\kwave_p = 33.5 \text{ m}^{-1}$)
  (b) $\flucu=0.36 \text{ ms}^{-1}$, $f_p = 70$~Hz, ($\kwave_p=58.6
  \text{ m}^{-1}$). 
  (c) $\flucu = 0.16 \text{ ms}^{-1}$, $f_p=70$ Hz,
  ($\kwave_p=58.6 \text{ m}^{-1}$).}
\label{fig.response1}
\end{figure}

%
The response $R(\kwave,\tau)=\left| \langle u(\kwave,\tau) \:
F^*(\kwave,0) \rangle \right|$
\footnote{Also known as the coherence spectrum.}
, with $k = 2\pi \: f_p / U$, is shown in Fig.\ \ref{fig.response1}
for various combinations of the forcing frequency that determines the
forced scale size (forced wave number), and the background turbulence
intensity together with the predictions of (Eq.\ \ref{eq.dia}). 
Long averaging times of $\approx 10^4$ large--eddy turnover times
were necessary to detect the correlation with the periodic modulation
in the turbulence signal.  The signals were divided in 375 blocks of
$2^{17}$ samples, and the (complex) quantity $u(\kwave, \tau)\:
F^*(\kwave, \tau)$ was computed for each of the blocks with its
fluctuations determining the error in the response $\left\langle |
u(\kwave, \tau)\: F^*(\kwave, \tau)| \right\rangle$; those
statistical errors are indicated by the error bars in Fig.\
\ref{fig.response1}.
We find that for large turbulent velocities, the response roughly
follows (Eq.\ \ref{eq.dia}) and decays faster with increasing wave
number.  However, similarly to the case where we induced the
perturbation with the active grid (\S\ \ref{sec.active.per}), the
decay of the response does not depend on the background turbulence
level; more than halving it does not result in a slower decay.  
On a semi--log plot, (Eq.\ \ref{eq.dia}) is a parabola. Surprisingly,
the decay of the response with delay time that we find is closer to
exponential, in agreement with \cite{kellogg.1980}.

%
\section{Conclusion} 
Studying the response of a turbulent flow on additional perturbations
is a very challenging experiment. The perturbations should not alter
the pre-existing background turbulence and and they must be
detectable to measure a response function. In addition to these, the
frequency (wave number) of the perturbation should be tunable.
Finally, long integration times are required to obtain an adequate
signal to noise ratio. 

The linear response of turbulence to perturbation is a controversial
issue. Several experimental attempts to measure this response have
been reported. An important guidance principle is Kraichnan's
prediction (Eq.\ \ref{eq.dia}) which assumes modulation at small
scales, which is scrambled by the random large scales. We have
imposed both the background turbulence and the modulation by the same
active grid. The drawback of these experiments is the poor separation
between the modulation and background turbulence and the relatively
small wave numbers. Indeed, the measured decay of the linear response
is slow in comparison to (Eq.\ \ref{eq.dia}), while the response does
not depend on the background turbulence level.

A much better separation between background turbulence and modulation
was achieved with a synthetic jet. Our central results (in Fig.\
\ref{fig.response1}) show that modulations of higher wave numbers
decay faster, in agreement with (Eq.\ \ref{eq.dia}), but that a
decrease of the background turbulence does not lead to a slower
decay, contrary to the prediction of (Eq.\ \ref{eq.dia}).  In this
respect, no other experiment has shown convincing support for (Eq.\
\ref{eq.dia}), neither has ours. 

\begin{acknowledgments}
This work is part of the research programme of the `Stichting voor
Fundamenteel Onderzoek der Materie (FOM)', which is financially
supported by the `Nederlandse Organisatie voor Wetenschappelijk
Onderzoek (NWO)'.  This work is also supported by the COST Action
MP0806.  We thank Gerald Oerlemans for technical assistance.
\end{acknowledgments}

\setlength{\bibsep}{1ex}
\setlength{\bibhang}{2ex}
\bibpunct{(}{)}{;}{a}{ }{,}
%

%
\end{document}